# Giant anomalous Hall effect from spin-chirality scattering in a chiral magnet

Yukako Fujishiro[1*†], Naoya Kanazawa[1*†], Ryosuke Kurihara[2], Hiroaki Ishizuka[1], Tomohiro Hori[1], Fehmi Sami Yasin[3], Xiuzhen Yu[3], Atsushi Tsukazaki[4], Masakazu Ichikawa[1], Masashi Kawasaki[1,3], Naoto Nagaosa[1,3], Masashi Tokunaga[2], and Yoshinori Tokura[1,3,5*]

[1] *Department of Applied Physics, The University of Tokyo, Bunkyo-ku, Tokyo 113-8656, Japan*
[2] *The Institute for Solid State Physics (ISSP), The University of Tokyo, Kashiwa, Chiba 277-8581, Japan*
[3] *RIKEN Center for Emergent Matter Science (CEMS), Wako, Saitama 351-0198, Japan*
[4] *The Institute for Materials Research (IMR), Tohoku University, Aoba-ku, Sendai 980-8577, Japan*
[5] *Tokyo College, The University of Tokyo, Bunkyo-ku, Tokyo 113-8656, Japan*

\* To whom correspondence should be addressed.

† These authors equally contributed to this work.

E-mail: fujishiro@cmr.t.u-tokyo.ac.jp, kanazawa@ap.t.u-tokyo.ac.jp, and tokura@riken.jp

**The electrical Hall effect can be significantly enhanced through the interplay of the conduction electrons with magnetism, which is known as the anomalous Hall effect (AHE). Whereas the mechanism related to band topology has been intensively studied towards energy efficient electronics, those related to electron scattering have received limited attention. Here we report the observation of giant AHE of electron-scattering origin in a chiral magnet MnGe thin film. The Hall conductivity and Hall angle respectively reach $40,000$ $\Omega^{-1}$cm$^{-1}$ and $18$ % in the ferromagnetic region, exceeding the conventional limits of AHE of intrinsic and extrinsic origins, respectively. A possible origin of the large AHE is attributed to a new type of skew-scattering via thermally-excited spin-clusters with scalar spin chirality, which is corroborated by the temperature-magnetic-field profile of the AHE being sensitive to the film-thickness or magneto-crystalline anisotropy. Our results may open up a new platform to explore giant AHE responses in various systems, including frustrated magnets and thin-film heterostructures.**

The quantum nature of electrons leads to rich electromagnetic responses in correlated electron systems, in particular, through the interaction with magnetism[1]. The anomalous Hall effect (AHE) is one such phenomenon, where the Hall effect occurs usually as a consequence of magnetic ordering and spin-orbit coupling (SOC)[2]. This phenomenon potentially benefits applications as well, since a large Hall response provides a pathway to energy efficient electronic or spintronic devices through the suppression of the longitudinal current which entails the Joule heating. To date, studies aiming at a large Hall response have focused on the gauge field arising from the electronic bands and magnetic orders. For example, a small magnetization can produce a large net Berry curvature in topological semimetals[3-7]. The extreme example is the quantized AHE in magnetic topological insulators, which realizes the infinite Hall angle and hence dissipationless conduction[8,9]. Another direction is to tailor a non-coplanar magnetic order with a geometrical correlation as characterized by scalar spin chirality (SSC) $\chi = \mathbf{S}_i \cdot (\mathbf{S}_j \times \mathbf{S}_k)$, which results in a large net Berry curvature through the coupling between conduction electrons[10-13]. Here, $\mathbf{S}_i$, $\mathbf{S}_j$, and $\mathbf{S}_k$ represent spins at three neighboring atomic sites $(i, j, k)$. In particular, a dense topological spin texture with a short magnetic period leads to large topological Hall effect (THE)[14,15]. However, the Hall conductivity of these "intrinsic" mechanisms has the upper threshold set by the Berry curvature. In the case of momentum space Berry curvature, the consequent Hall conductivity should be less than $e^2/ha$ ($h$ and $a$ being Planck's constant and a typical lattice constant values), and hence of the order of $\sigma_{xy} = 10^2$-$10^3$ $\Omega^{-1}$cm$^{-1}$.

On the other hand, the contribution of electron scattering (conventionally termed "extrinsic" mechanism) is not restricted by the Berry curvature. Nevertheless, extrinsic mechanisms have rarely been studied in the context of the large AHE responses, due to its small Hall conductivity $(\sigma_{xy})$. The only exception occurs in the extremely high conductivity regimes (empirically above $\sigma_{xx} > 5 \times 10^5$ $\Omega^{-1}$cm$^{-1}$), where the skew-scattering (asymmetric electron scattering due to the SOC at impurities) dominates $\sigma_{xy}$ with a characteristic scaling relation $(\sigma_{xy} \propto \sigma_{xx})$[16,17]. Even in that case, the Hall angle $[\theta_H = \tan^{-1}(\sigma_{xy}/\sigma_{xx})]$ remains a constant value, i.e., $\sigma_{xy} \propto \sigma_{xx}$, which is as small as 0.1– 1 % because the spin-orbit interaction is usually smaller than the width of the virtual bound state and also skew scattering requires the phase shifts of different orbital angular momenta[16-21].

**Results**

**MnGe thin film with topological spin texture**

Here we report the observation of a giant AHE in a chiral magnet MnGe thin film, where the giant Hall angle (18 %) leads to a Hall conductivity reaching $\sigma_{xy} \sim 40,000$ $\Omega^{-1}$cm$^{-1}$, being two orders of magnitude larger than the intrinsic AHE, even with a moderate longitudinal conductivity ($\sigma_{xx} \sim 2 \times 10^5$ $\Omega^{-1}$cm$^{-1}$). The target material is the epitaxially grown thin films of MnGe[22], a member of the B20-type chiral magnets hosting topological spin textures[23,24]. The crystal structure belongs to the non-centrosymmetric space group $P2_13$, where the lattice chirality is characterized by the stacking direction of the atoms as viewed from [111] axis (only one enantiomeric form is shown in Fig. 1**a**). Since the lack of inversion symmetry allows Dzyaloshinskii-Moriya interaction (DMI), twisted spin structures such as helical or non-coplanar structures are often observed in B20-type magnets[23,24]. Among them, MnGe hosts unique magnetic textures composed of spin hedgehogs and anti-hedgehogs. In a bulk sample, those magnetic defects form a short-period lattice (~ 2.8 nm) with large magnetic fluctuations even persisting in the nominally ferromagnetic region[24]. The hedgehog lattice is deformed to a rhombohedral form (Fig. S1)[22] or decomposed into multi-domains with differently-oriented helical structures[25] in strained thin films (see Supplementary Note 1 for comment on the magnetic texture in thin films). In either case, there exist chains of hedgehogs and anti-hedgehogs connected by skyrmion strings. Such a non-coplanar magnetic order produces a THE in metallic systems including MnGe ($\sigma_{xx} = 2.0 \times 10^5$ $\Omega^{-1}$cm$^{-1}$ at 2 K) (Fig. 1**b**); the large negative THE below $T < 50$ K (Fig. 1**c**) is ascribed to emergent magnetic field from the skyrmion-strings (inset of Fig. 1**c**)[14].

**Hall conductivity in the ferromagnetic region**

The main focus of this work is on the high-magnetic-field Hall response shown in Fig. 1**d**, where $\sigma_{xy}$ shows a striking enhancement reaching ~40,000 $\Omega^{-1}$cm$^{-1}$ at the temperature $T = 2$ K, with a sharp peak structure at around the magnetic field $B = 14$ T (see Fig. S2 for the resistivity data). Whereas $\sigma_{xx} = 2.02 \times 10^5$ $\Omega^{-1}$cm$^{-1}$ at 2 K (Fig. 1**b**) belongs to the empirical intrinsic regime where the Berry curvature mechanism is dominant, the observed $\sigma_{xy}$ far exceeds the threshold value of the intrinsic AHE from the momentum space, which is roughly estimated from the quantization limit ($\sim e^2/ha = 800$ $\Omega^{-1}$cm$^{-1}$ for MnGe) in three dimensions (denoted by the dashed line in Fig. 1**d**)[17]. Here, $a = 4.795$ Å for MnGe. Upon increasing the temperature, $\sigma_{xy}$ is rapidly suppressed while showing a broader peak structure. Above $T \sim 70$ K, $\sigma_{xy}$ follows the conventional behavior of the intrinsic AHE, which scales with the magnetization[2] (Fig. S2).

**Scaling relation and temperature-magnetic-field profile of the Hall response**

To understand the characteristics of this large Hall response observed at low temperatures, we first plotted the data of $\sigma_{xy}$ (Fig. 1**d**) against $\sigma_{xx}$ (Fig. S2) to investigate their scaling relation, which has been typically used to identify the mechanism of the AHE[2,16,17]. Figure 2**a** shows the plot for various samples with different film-thicknesses ($t = 80, 160,$ and $300$ nm) for $T$ = 2-30 K (see Fig. S3 for the complete data set). The displayed data points ($\sigma_{xx}$, $\sigma_{xy}$) are those exhibiting the maximum Hall angle at each temperature. In every sample, the scaling relation between $\sigma_{xy}$ and $\sigma_{xx}$ is linear ($\sigma_{xy} \propto \sigma_{xx}$), which is a typical feature expected for skew-scattering mechanism (see Fig. S4 and Supplementary Note 4 for detailed discussions). To overview this Hall response in the $B$-$T$ plane, the contour map of $\sigma_{xy}$ is shown with the magnetic phase diagram in Fig. 2**b**. The striking enhancement of the Hall response appears at low temperatures below 50 K, in proximity to the ferromagnetic (FM) phase boundary. The constant-$B$ cut of $\sigma_{xy}$ shows a peak structure at finite temperature, and the peak structure shifts to higher temperatures under higher magnetic fields (Fig. 2**c**).

The above features of the skew-scattering in MnGe are distinct from the conventional skew-scattering induced by the non-magnetic chemical defects or single-spin impurity[2,16-21], while showing the similar linear ($\sigma_{xy} \propto \sigma_{xx}$) scaling relation. In the conventional skew-scattering, the perturbatively small SOC compared to the electron bandwidth results in a small Hall angle of 0.1– 1 %. Also, $\sigma_{xy}$ should be monotonously increased to lower temperature with higher $\sigma_{xx}$. Hence, not only the giant Hall angle of 18~22 % (Fig. 2**a**) but also the non-monotonous $B$-$T$ profile (Fig. 2**b,c**) are both beyond the conventional understanding of skew-scattering. One possibility for this unconventional skew-scattering is the recently proposed "spin-chirality skew-scattering" mechanism[26], where the thermally fluctuating spins in the ferromagnetic state act as the spin-clusters with the SSC (inset of Fig. 2**c**). Hence, the magnetic scattering contributes to the AHE. In principle, the scattering process there does not involve SOC and the Hall angle is predicted to be significantly large compared to the conventional skew-scattering. In particular, the Hall angle in the order of 10 % is anticipated to show up when the size of the spin-cluster is comparable to the inverse of the Fermi-wave-vector ($\sim\frac{1}{k_F}$), which corresponds to the resonance condition of the interference effect[26]. Moreover, the observed $B$-$T$ profile of the AHE is also consistent with the spin-cluster scenario. Under fixed magnetic field, the spin-cluster AHE is expected to be maximized at a finite temperature as shown in Fig. 2**c**. Two types of spin excitations with opposite SSC are thermally excited, being responsible for skew scatterings with opposite Hall angles. Because those SSC excitations have different activation energies due to the presence of DMI, the cancellation of AHE signal only occurs above the temperature where the SSC excitations with the higher activation energy start to proliferate. The overall $B$-$T$ profile of the AHE is well reproduced by analytical calculation in terms of the thermal excitation of the SSC (see Fig. S8 and Supplementary Note 6 for detail).

**Film-thickness dependence of the Hall response**

To elaborate on the possibility of the spin-cluster mechanism, we have controlled the magneto-crystalline anisotropy by changing the film-thickness; the SANS experiment on MnGe thin films has revealed the enhanced in-plane magnetic anisotropy with decreasing film-thickness[22] (Fig. S1). Therefore, the spins can "tilt" easier from the field-polarized direction in the FM state in thinner films and we expect that the SSC excitation in the collinear spin background has a corresponding lower energy. In other words, larger SSC can be produced due to the enhanced in-plane anisotropy in the thinner films as schematically illustrated in the insets of Fig. 3. We note that the insets of Fig. 3 are overdrawn to emphasize the difference induced by the film-thickness. The $B$-$T$ profile of the AHE shows a clear variation with film-thickness of $t = 80, 160$, and 300 nm (Fig. 3, see also Figs. S2 and S3 for the complete data set). While the maximum value of the Hall angle is almost independent of the film-thickness, the $B$-$T$ profile of the AHE changes dramatically as shown in Fig. 3. In thinner films with enhanced in-plane anisotropy, a large Hall angle shows up at lower temperature and higher applied magnetic field, where the SSC excitation cost more energy. We also provide analytical calculations on the effect of easy-plane anisotropy in Supplementary Note 6.

**Comparison with other materials**

The discovery of this new type of the AHE, showing giant Hall conductivity and Hall angle simultaneously, provides a distinct exception in the universal scaling curve of $\sigma_{xy}$ versus $\sigma_{xx}$ established for various ferromagnets[2,17] (Fig. 4). The Hall angle of 18 % in MnGe gives one- or two-orders of magnitude larger $\sigma_{xy}$ for a given $\sigma_{xx}$, compared to the conventional skew-scattering, resulting in an upward shift of the scaling plot with a linear relation ($\sigma_{xy} \propto \sigma_{xx}$). We also emphasize that $\sigma_{xy}$ from the intrinsic AHE cannot reach this regime, due to the limitations set by the Berry curvature in momentum space. Recently, the spin-cluster AHE was also reported in a frustrated magnet KV$_3$Sb$_5$, with large Hall conductivity of 15,507 $\Omega^{-1}$cm$^{-1}$ and Hall angle of 1.8 %[27]. We assume that the difference in the Hall angle between MnGe and KV$_3$Sb$_5$ may be related to the resonance condition for the spin-chirality skew-scattering. For MnGe, the unusually short magnetic period ($\lambda \sim 2.8$ nm) has been discussed in terms of the conduction-electron mediated mechanisms[24,28,29] such as the Rudermann-Kittel-Kasuya-Yosida (RKKY) interaction; this can automatically tune the size of the spin cluster to the typical size of $\sim \frac{1}{k_F}$, satisfying the resonance condition. Hence the RKKY magnets or the metallic spin-glass systems would be the promising candidates for achieving a large Hall angle. However, the true nature of the excitation modes for the SSC in MnGe remains to be a challenge for future study. For instance, the origin of the SSC can be either the slightly tilted spin clusters or the pairwise excitations of spin hedgehogs and anti-hedgehogs connected by the skyrmion strings. We speculate that the latter may be the case, given the fact that the magnetic ground state of MnGe at low temperature ($T < 50$ K) is an ordered state of these spin singularities.

**Discussions**

One another possibility for the large Hall response observed in MnGe films is the emergence of high-mobility carriers in the ferromagnetic region, which can result in $\sigma_{xy}$ with a sharp peak structure, as typically observed in Dirac or Weyl semimetals[30]. We found that the observed Hall conductivity could be roughly reproduced by two-carrier Dude model, by assuming a presence of high-mobility ($\sim$690 cm$^2$V$^{-1}$cm$^{-1}$) and low carrier-density ($\sim$9.4 $\times$ 10$^{20}$ cm$^{-3}$) hole pocket (see Supplementary Note 5 and Figs. S5-S6 for detailed discussions). However, this assumption that the observed Hall response is dominated by the normal Hall effect is less plausible, partly because the giant Hall response is not observed in bulk polycrystalline MnGe which has the almost same $\sigma_{xx}$ (1.6 $\times$ 10$^5$ $\Omega^{-1}$cm$^{-1}$ at 2 K) with that of the thin film, but the SSC excitation effect may be cancelled out due to the randomly-oriented crystalline domains (see Supplementary Note 5 and Figs. S7 for detailed discussions). Further experiments such as the direct observation of electronic structure, especially in the ferromagnetic region, would be necessary to discuss the possibility of large normal Hall effect or the presence of magnetic Weyl points in MnGe. Moreover, the characteristic features discussed in this work, such as the scaling relations, non-monotonous *B-T* profile, and magneto-crystalline anisotropy dependence of the Hall response, strongly suggest the spin-chirality skew-scattering mechanism argued above.

In conclusion, we have observed giant AHE in the moderately conductive MnGe thin films, which suggests the skew-scattering from the scalar spin chirality (SSC) excitation. Since most of the experimentally available materials belong to the intrinsic regime ($10^3 < \sigma_{xx} < 10^5$) or the dirty regime ($\sigma_{xx} < 10^3$), spin-chirality skew-scattering would provide an opportunity to explore giant AHE responses in various materials. It is expected to show up in systems where the finite thermal average of the SSC is induced by the short-range spin correlation, for instance by the geometrical frustration or the inversion-symmetry breaking at the interface. In that sense, the large family of frustrated magnets[31] with triangular, Kagome, and pyrochlore lattices, as well as the heterostructures with interfacial DMI[32] would be prospective platforms to explore this new AHE. Unlike the conventional intrinsic AHE which has been limited to ferromagnets or the ordered phases of the SSC, the proposed new paradigm of spin-chirality skew-scattering would be applicable to a wide range of materials; since it is expected even in materials with no magnetic-ordering as well as in the temperature-magnetic region where the SSC loses its long-range order. Exploration of giant AHE responses, from a perspective of short-range spin correlation, would open up a new frontier in the discovery of novel electronic functionalities.

**Methods**

**Thin-film growth**

MnGe thin films were grown epitaxially on Si(111) substrates by the molecular beam epitaxy method[22]. We employed a 2nm-thick MnSi(111) buffer layer, which was grown by reacting a deposited Mn layer with the Si(111)-(7×7) surface at 250 ℃. Then, Mn and Ge were co-evaporated at approximately 90 ℃, followed by annealing at 250 ℃. The growth of B20-type MnGe was confirmed by $\theta$-$2\theta$ x-ray diffraction scans.

**Electric transport measurement**

Resistivities up to 14 T were measured with a conventional four-probe method using a DC option of Physical Properties Measurement System (PPMS, Quantum Design). High-magnetic-field transport measurements up to 56 T were performed utilizing non-destructive pulse magnets energized by capacitor banks installed at the International MegaGauss Science Laboratory of Institute for Solid State Physics (ISSP), University of Tokyo. The magnetic field was applied perpendicular to the film plane, *i.e.*, (111) plane, while the electric field was applied parallel to $[1\bar{1}0]$ direction. The longitudinal conductivity ($\sigma_{xx}$) and the Hall conductivity ($\sigma_{xy}$) were calculated as $\sigma_{xx} = \rho_{xx}/(\rho_{xx}^2 + \rho_{yx}^2)$ and $\sigma_{xy} = \rho_{yx}/(\rho_{xx}^2 + \rho_{yx}^2)$. Here, $\rho_{xx}$ and $\rho_{yx}$ are the longitudinal and Hall resistivity, respectively. We could not perform the transport measurements in thick samples (film-thickness larger than 400 nm) due to the formation of microcracks upon cooling.

**Dark-field transmission electron microscopy (TEM) measurement**

The epitaxially-grown MBE thin films of MnGe are composed of a mixed-domain state of enantiomers, although the different enantiomer (lattice chirality) with an opposite-sign Dzyaloshinskii-Moriya interaction should show an identical contribution to any kind of Hall effect. The size of the chiral domain in MnGe thin films was investigated by the dark-field TEM measurement[36,37]. First, we prepared the sample by grinding the Si substrate down to ∼ 6-8 μm using the multiprep and dimple grinding systems. We then utilized precision ion milling (Pips) to mill the remaining Si substrate, leaving the 80 nm, 160 nm and 300 nm thick films, which were mounted onto TEM sample holder-ready Mo annuli. We characterized the orientation of each film using electron diffraction. As in Ref. [22], we found the [111] axis incidence using selected area electron diffraction within a JEOL JEM 2100 TEM as shown in Fig. S9**a**. We utilized a LN-holder (Gatan 636) for its high angle tilt capability. As shown in Figs. S9**a, b**, the diffraction peaks corresponding to the two domains of interest (labelled A and B in Fig. S9**b**) emerge when the crystal is tilted around the [11$\bar{2}$] axis by approximately 22º. The spots were isolated using the objective aperture, and the passing electrons were focused into an image of the domains onto the charge-coupled device (CCD) camera installed within the microscope.

**Data availability**

The data sets generated during and/or analysed during the current study are available from the corresponding author on reasonable request.

**Acknowledgment**

The authors thank A. Kitaori and M. Mogi for experimental supports and fruitful discussions. This research was supported in part by JSPS KAKENHI (Grants No. JP18K13497, No. JP20H01859, and No. JP20H05155) and JST CREST (Grants No. JPMJCR16F1 and No. JPMJCR1874).

**Author contributions**

Y.T. and N.K. conceived the project. Transport and magnetization measurements using PPMS/MPMS were performed by Y.F. and N.K.. High-magnetic-field transport measurements were performed by Y.F., N.K., R.K. M.T. Thin films were grown by N.K. and T.H.. Theoretical calculations were provided by H.I. and N.N.. Dark-field TEM measurements were performed by F.S.Y. and X.Z.Y.. Y.F., N.K., and Y.T. wrote the paper with the support from H.I., A.T., M.I., M.K., N.N., M.T..

**Competing interests**

The authors declare no competing interests.

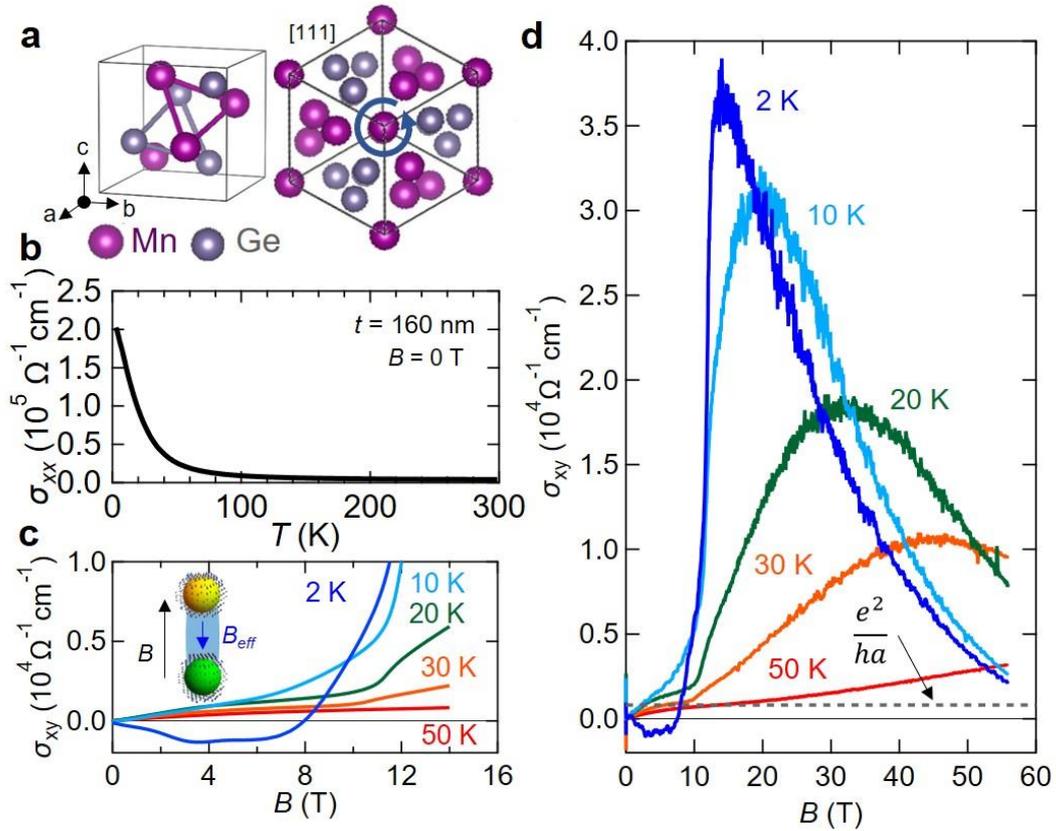

**Figure 1 | Observation of the giant anomalous Hall effect (AHE) in MnGe thin film. a,** B20-type chiral crystal structure (space group $P2_13$) of MnGe. The chirality is characterized by the stacking direction of atoms as viewed from [111] direction. **b,** Temperature dependence of conductivity ($\sigma_{xx}$) at zero magnetic field for the film-thickness of $t = 160$ nm. **c,** Magnetic-field dependence of Hall conductivity ($\sigma_{xy}$) at various temperatures for $t = 160$ nm. The negative dip structures are attributed to topological Hall effect arising from the formation of spin hedgehogs and anti-hedgehogs bridged by skyrmion-strings (shown in the inset). **d,** High-magnetic-field data of $\sigma_{xy}$ at various temperatures for $t = 160$ nm, showing a large enhancement. The value of the

quantization limit ($\frac{e^2}{ha}$) in three-dimensions is denoted by a dashed line. The observed giant AHE far exceeds the value allowed by the intrinsic AHE.

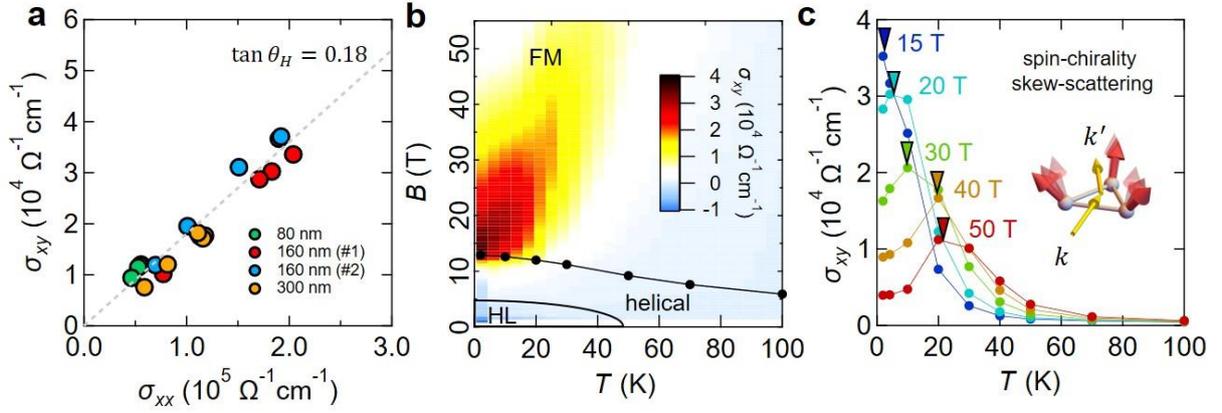

**Figure 2 | Characteristics of the AHE observed in MnGe thin film, suggesting a spin-chirality skew-scattering. a,** Plot of Hall conductivity ($\sigma_{xy}$) versus conductivity ($\sigma_{xx}$) for the thickness of $t$ = 80, 160, and 300 nm. The data are taken from the peak positions of the Hall angle below 30 K. The linear relation ($\sigma_{xy} \propto \sigma_{xx}$) appears in a certain $B$-$T$ region with a large Hall angle of ~ 18 %. **b,** Contour plot of $\sigma_{xy}$ in the $B$-$T$ space with the magnetic phase diagram consisting of the hedgehog lattice (HL), helical, and the ferromagnetic (FM) phases. **c,** Constant-$B$ cut of $\sigma_{xy}$ showing a peak structure at a finite temperature. The triangles denote the peak positions, which shifts to a higher temperature at a higher magnetic field. These results are suggestive of the relation to the thermal excitation of the scalar spin chirality (inset of Fig. 2**c**).

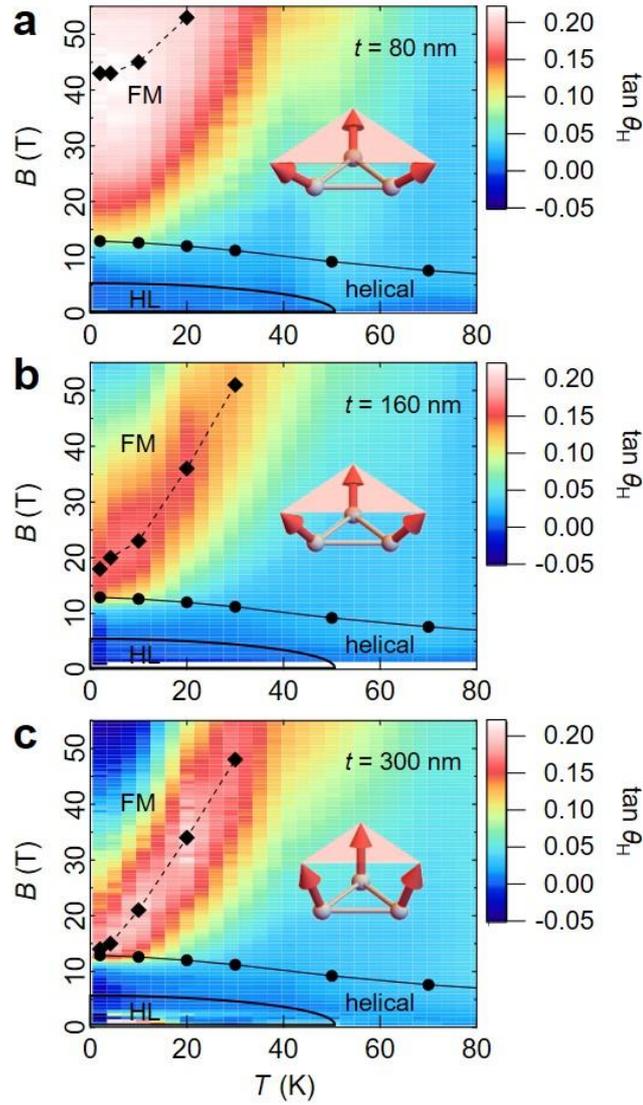

**Figure 3 | Film-thickness dependence of the temperature-magnetic field profile of the AHE in MnGe thin film.** Contour plot of $\tan\Theta_H$ ($= \sigma_{xy}/\sigma_{xx}$) for the thickness of $t = 80$ (**a**), $160$ (**b**), and $300$ nm (**c**). The solid diamond markers connected by a dashed line represent the maximum point of $\tan\theta_H$ at each temperature. The insets are the intuitive schematics to explain the fact that the larger scalar spin chirality excitation is more favourable in thinner films with enhanced in-plane magnetic anisotropy.

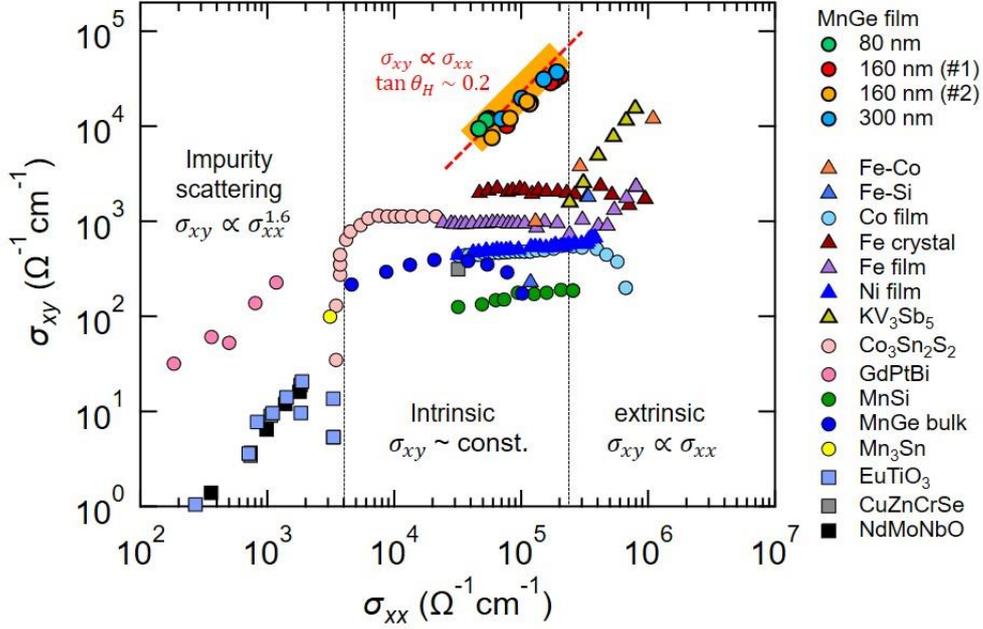

**Figure 4 | Full logarithmic plot of $\sigma_{xy}$ versus $\sigma_{xx}$ for various materials.** The shaded region with a red dashed line indicates a linear relation ($\sigma_{xy} \propto \sigma_{xx}$) in MnGe thin films with a giant Hall angle reaching $18 \sim 22\%$ for the thickness of $t = 80, 160$, and 300 nm. The data are taken from the peak positions of the Hall angle below 30 K. The large Hall conductivity and the Hall angle realized in MnGe thin films result in a clear deviation from the conventional scaling plots established for ferromagnets[2,18,19]. The reported data for other materials are cited from literature[6,7,14,16,17,20,27,33-35].